\documentstyle[11pt]{article}
\begin{document}

\newcommand{\vm}{\vspace{0.2cm}}
\newcommand{\vl}{\vspace{0.4cm}}

\vspace{0.2cm}
\begin{center}
\begin{Large}
{\bf  Color Confinement and Massive Gluons}
\end{Large}
\vskip 10mm {{\bf Masud Chaichian$^1$ and Kazuhiko
Nishijima$^{1,2}$}}

\vspace{0.2cm}
      {\small   $^1$Department of Physics, High Energy Physics Division}\\
           {\small University of Helsinki} \\
{\small and}\\ {\small Helsinki Institute of Physics}\\ {\small P.O.
Box 9, FIN-00014 Helsinki, Finland}\\ {\small $^2$Department of
Physics, University of Tokyo, 7-3-1 Hongo,\\ Bunkyo-ku, Tokyo
113-0033, Japan}
\date{}
\end{center}


\vspace{2.0cm}
\begin{center}\begin{minipage}{5in}
\begin{center} {\bf Abstract} \end{center}
\baselineskip 0.3in{\small Color confinement is one of the central
issues in QCD so that there are various interpretations of this
feature. In this paper we have adopted the interpretation that
colored particles are not subject to observation just because
colored states are unphysical in the sense of Eq. (2.16). It is
shown that there are two phases in QCD distinguished by different
choices of the gauge parameter. In one phase, called the
"confinement phase", color confinement is realized and gluons turn
out to be massive. In the other phase, called the "deconfinement
phase", color confinement is not realized, but the gluons remain
massless.
}\\

\vskip 10mm
\end{minipage}
\end{center}


\section{Introduction}

Strong interactions of hadrons are considered to be
governed by QCD dealing with the interactions of
quarks and gluons. It is a gauge theory based on
the color $SU(3)$ group and a strongly interacting
particle belongs to an irreducible representation
of this group, for instance, quarks and gluons
belong to the color triplet and octet
representations, respectively. It so happens,
however, that those particles that belong to
non-singlet representations are not subject to
direct observations and this property is referred
to as color confinement. Thus both quarks and
gluons are unobservable, and the only observable
particles are color singlet composite particles
called hadrons.

Experimentally, we can study the properties of
quarks and gluons only through hadrons. In order to
study hadron-hadron interactions theoretically we
shall first illustrate the problem by a similar one
in QED. The interaction between two charged
particles is governed by Coulomb's law, but the
interaction between two electrically neutral
particles is represented by the van der Waals
potential,

$$\begin{array}{l} V_{vdW}(r)\propto r^{-6}  .
\end{array}\eqno{(1.1)}$$

\noindent This shows that the electric fields
generated by neutral systems penetrate into the
vacuum without any sharp cut-off.

In QCD we can elucidate the dynamical properties of
color-neutral hadrons with reference to dispersion
relations. For the scattering of hadrons, for
instance, they remain applicable provided that
confinement excludes quarks and gluons from the
physical intermediate states appearing in the
unitarity conditions. This is precisely the
condition for color confinement. As a typical
example, let us consider nucleon-nucleon
scattering, then the potential between them is
given by the pole contribution in the crossed
channels. The least massive hadron that can be
exchanged between them is the pion, and the
resulting interaction is represented by the Yukawa
potential,

$$\begin{array}{l} V_{Y}(r)\propto \frac{exp(-\mu
r)}{r}  ,
\end{array}\eqno{(1.2)}$$

\noindent where $\mu$ denotes the pion mass.

In comparison with the van der Waals force in QED
we recognize that the flux of  the color gauge
field emerging from color singlet nucleons cannot
penetrate into the confining vacuum beyond a
certain range thereby leaving no trace of
long-range forces and that the penetration depth is
given by the pion Compton wave length. Thus we
recognize a similarity between the Yukawa mechanism
for generating nuclear forces and the Meissner
effect in the type II super conductors. This
similarity strongly suggests that the vacuum allows
penetration of the flux of the color gauge field
generated by hadrons only by a finite length.

Thus we many point out two salient features of QCD
on the basis of the experimental properties of
hadrons: i) color confinement, ii) finite
penetration length of the fluxes of the color gauge
field.

The first one, color confinement, has been elucidated in
\cite{1}-\cite{4} and we shall concentrate ourselves on the second
problem in the present article. Actually, we interpret ii), finite
penetration length, as an evidence for massive gluons and we shall
show that realization of color confinement as interpreted in
\cite{1}-\cite{4} automatically leads to massive gluons. This is the
main subject of this paper.

In Sec. 2 we shall recapitulate the arguments for color confinement
based on BRS invariance \cite{7} and asymptotic freedom \cite{5,6}.
In Sec. 3 we shall show that gluons turn out to be massive when
color confinement is realized. In Sec. 4 we show that confinement is
realized only in the gauges in which $Z_3^{-1}=0$ is realized. This
is referred to as the confinement phase. On the other hand, gluons
remain massless in the other gauges in which the above condition is
not met, and this case is referred to as the deconfinement phase.

\section{BRS Invariance and Color Confinement}

In this section we shall recapitulate the essence of the
interpretation of color confinement that has been developed in a
series of articles \cite{1}-\cite{4}.

In a covariant quantization of gauge fields introduction of
indefinite metric is indispensable. Thus the resulting state vector
space ${\cal V}$ involves unphysical states of indefinite metric and
we have to find a criterion to select physical states out of ${\cal
V}$.  For this purpose we employ the Lorentz condition in QED as a
subsidiary condition, but it is more involved in non-Abelian gauge
theories. In what follows we shall confine ourselves to QCD, and in
order to fix the notation we start from its Lagrangian density in
the metric $g_{\mu\nu}=diag(1,-1,-1,-1)$

$$
\begin{array}{lll}
{\cal L} = {\cal L}_{inv}+ {\cal L}_{gf}+{\cal
L}_{FP},
\end{array}
\eqno {(2.1)} $$ \noindent where
 $$\begin{array}{lll}
\hspace{1.0cm}&     {\cal L}_{inv}= -\frac{1}{4}F^{\mu\nu}\cdot
F_{\mu\nu} +i \overline{\psi}(\gamma^{\mu} D_{\mu}-m)\psi , &
\hspace{6.0cm} (2.2a) \\ &\\ \hspace{1.0cm}& {\cal
L}_{gf}=-A^{\mu}\cdot \partial_{\mu} B +\frac{\alpha}{2} B\cdot B ,
& \hspace{6.0cm} (2.2b)\\ &&\\ \hspace{1.0cm}& {\cal L}_{FP}=
-i\partial^{\mu} \overline{c} \cdot D_{\mu}c   &
\hspace{6.0cm}(2.2c)\\
\end{array}
$$

\noindent in the customary notation \cite{1}-\cite{4}.   We have
suppressed the color and flavor indices in (2.2).  The second
Lagrangian density (2.2b) is the gauge-fixing term in which
 $\alpha$ denotes the gauge parameter and
$B$ the Nakanishi-Lautrup auxiliary field. The last one (2.2c) is
the Faddeev-Popov ghost  term, and the hermitian scalar fields $c$
and $\bar{c}$ are anticommuting and are called Faddeev-Popov (FP)
ghost fields.  The local gauge invariance is respected only by the
first term (2.2a) but not by the other two, (2.2b) and (2.2c),
introduced for the purpose of quantization. The total Lagrangian is
invariant, however, under the global BRS transformations \cite{7}
defined below.

\vm {\bf BRS transformations} \vm

 Let us consider
an infinitesimal gauge transformation of the gauge
and quark fields and replace the infinitesimal
gauge function either by $c$ or $\overline{c}$.
They define two kinds of BRS transformations
denoted by $\delta$ and $\overline{\delta}$,
respectively.

$$
\begin{array}{lll}
\delta A_{\mu}= D_{\mu}c , & \overline{\delta}
A_{\mu}=
D_{\mu}\overline{c} ,\\
\end{array}
\eqno{(2.3)}
$$
$$
\begin{array}{lll}
\delta \psi= ig (c\cdot T) \psi ,&
\overline{\delta}  \psi= ig
(\overline{c}\cdot T) \psi .\\
\end{array}
\eqno{(2.4)}
$$
\noindent where the matrix $T$ is introduced in the
covariant derivative of $\psi$ as

$$\begin{array}{l}  D_{\mu} \psi = (\partial
_{\mu} -i g T \cdot A_{\mu} ) \psi, \\
\end{array}
\eqno{(2.5)}$$

\noindent For the auxiliary fields $B$, $c$ and
$\bar{c}$ local gauge transformations are not even
defined, but their BRS transformations can be
introduced by requiring the invariance of the local
Lagrangian density, namely,

$$
\begin{array}{lll}
\delta {\cal L} = \overline{\delta} {\cal L}= 0.
\end{array}
\eqno{(2.6)} $$ \noindent We shall not write them down explicitly,
however, since they are not relevant to the following arguments.
Noether's theorem states that the BRS invariance of the Lagrangian
density amounts to two conserved BRS charges denoted by $Q_B$ and
$\overline{Q}_B$. They satisfy

$$
\begin{array}{lll}
\delta \phi= i \left[ Q_B,\phi\right]_{\mp} , &
\overline{\delta} \phi= i \left[
\overline{Q}_B,\phi\right]_{\mp}    ,
\end{array}
\eqno{(2.7)}
$$
\noindent where we choose the $-(+)$ sign, when the
field $\phi$ is of an even (odd) power in the ghost
fields $c$ and $\overline{c}$.

Equations of motion for  the gauge field can be expressed with the
help of BRS transformations as \cite{1}-\cite{4}

$$
\begin{array}{lll}
\partial^{\mu}F_{\mu\nu} + gJ_{\nu}= i\delta\overline{\delta}
 A_{\nu} ,
\end{array}
\eqno{(2.8)}
$$
 \noindent where $J_{\nu}$ denotes the color current density and $g$ the
gauge coupling constant. It is worth noting that
all three terms in Eq. (2.8) are divergenceless
separately, in particular

$$
\begin{array}{lll}
\partial^{\nu}(i\delta\overline{\delta}
 A_{\nu}) =0.
\end{array}
\eqno{(2.9)} $$

 \noindent The BRS charges are
hermitian and nilpotent, for example,

$$
\begin{array}{lll}
Q_B^{\dagger}= Q_B , & Q_B^2=0 .\\
\end{array}
\eqno{(2.10)}
$$
\noindent The nilpotency implies introduction of
indefinite metric and a physical state $\vert
f\rangle$ is defined by the constraint

$$
\begin{array}{lll}
Q_B\vert f\rangle =0  , &  \vert f\rangle \in {\cal V}  .\\
\end{array}
\eqno{(2.11)}
$$
 \noindent The set of physical states including the vacuum state
$\vert 0\rangle$ forms the physical subspace of
${\cal V}$ denoted by ${\cal V}_{phys}$,

$$
\begin{array}{lll}
{\cal V}_{phys} =\{ \vert f\rangle: Q_B\vert
f\rangle =0  ,   \vert f\rangle \in {\cal V}\} .
\end{array}
\eqno{(2.12)} $$

\noindent Then the S matrix is BRS invariant and
satisfies

$$
\begin{array}{lll}
\delta S = i\left[ Q_B, S \right] =0  ,
\end{array}
\eqno{(2.13)} $$

\noindent so that the  physical subspace ${\cal
V}_{phys}$  is an invariant subspace of the S
matrix.

Furthermore, we introduce a subspace of ${\cal V}$
called the daughter subspace ${\cal V}_d$ defined
by

$$
\begin{array}{lll}
{\cal V}_{d} =\{ \vert f\rangle: \vert
f\rangle=Q_B\vert g\rangle
,   \vert g\rangle \in {\cal V}\}  .\\
 \end{array}
\eqno{(2.14)} $$

 \noindent Then because of the
nilpotency of $Q_B$, ${\cal V}_d$ is a subspace of
${\cal V}_{phys}$,

$$
\begin{array}{lll}
{\cal V}_{d} \subset {\cal V}_{phys},
 \end{array}
\eqno{(2.15)} $$

 \noindent and we introduce the
Hilbert space ${\cal H}$ by

$$
\begin{array}{lll}
{\cal H}= {\cal V}_{phys}/{\cal V}_{d}   .
 \end{array}
\eqno{(2.16)} $$

\noindent  We may compare ${\cal V}_{phys}$, ${\cal V}_{d}$ and
${\cal H}$ to closed forms, exact forms, and cohomology in Cartan's
algebra and (2.16) may be called the BRS cohomology \cite{3,9,10}.

When the quark  and gluon states are not physical,
they are unobservable and hence confined. Thus the
problem of color confinement reduces to
demonstration of the conditions

$$
\begin{array}{lll}
Q_B\vert quark\rangle \neq 0  , &  Q_B\vert gluon\rangle \neq 0  .\\
\end{array}
 \eqno{(2.17)}
$$

In a series of papers \cite{1}-\cite{4} it has been shown that the
criterion for color confinement takes a simple form

$$\begin{array}{l} C=0 ,\\
\end{array}\eqno{(2.18)}$$

\noindent where the constant $C$ is defined by

$$
\begin{array}{ll}
\partial^{\nu} \langle i \delta \bar{\delta} A_{\nu}^{a} (x),
A_{j}^{b} (y) \rangle = i \delta_{ab} C \partial
_{j} \delta ^4 (x-y) & (j=1,2,3).
\end{array}
\eqno{(2.19)}$$

\noindent Here and in what follows  $\langle
\cdots\rangle$ denotes the vacuum expectation value
of the time-ordered product.

Furthermore, with the help of the renormalization group and
asymptotic freedom \cite{1}-\cite{4}, \cite{10}, it has been shown
that the conditions (2.18) follows from

$$\begin{array}{l} Z_3^{-1}=0 , \end{array}\eqno{(2.20)}$$

\noindent where $Z_3$ is the renormalization
constant of the color gauge field.

It has already been shown by Oehme and Zimmermann \cite{11'} without
reference to perturbation theory that the above condition is
satisfied in the Landau gauge for $N_f<10$  on the basis of RGE and
asymptotic freedom. It is true that the anomalous dimensions in RGE
are evaluated in perturbation theory, but even then some results are
valid beyond perturbation theory thanks to asymptotic freedom.
  Depending on the nature of the objects RGE provides results at different
levels of precision, namely, (1) approximate or semi-perturbative,
or (2) exact. For instance, evaluation of the coefficient functions
in the operator product expansion falls into the first category,
whereas examination of some global properties of the theory falls
into the second category.  In fact, asymptotic freedom itself
follows from the negative beta function evaluated in the lowest
order, but this concept of asymptotic freedom is considered to be
valid beyond perturbation theory. Likewise, the condition (2.20) for
color confinement falls into the second category.
  This condition has been derived from unbroken non-abelian gauge symmetry
and asymptotic freedom related to the high energy behavior of the
quark-gluon system. When $Z_3^{-1}$ is expressed as the integral of
the Lehmann spectral function of the gluon propagator over the
entire energy region, however, Eq (2.20) indicates a delicate
balance of the contributions from all energy regions including both
ultraviolet and infrared. In this sense this condition gives an
indirect constraint on the infrared behavior of this system.  This
should be contrasted with Wilson's area law in which the low-energy
or long-distance behavior plays a dominant role in explaining quark
confinement. It is also worth emphasizing that as a consequence of
the above condition all the particles belonging to non-singlet
representations of the color group are confined for the same cause.

\section{Pole Structure of Green's Functions and
Massive Gluons}

In this section we shall show that gluons turn out
to be massive when the condition for color
confinement (2.18) is satisfied. For this purpose
we start form Eq. (2.8). This equation has been
given in the unrenormalized form so that we  put
the script (0) to unrenormalized expressions and
rewrite (2.8) as

$$
\begin{array}{lll}
\partial^{\mu}F^{(0)}_{\mu\nu} + g_0J^{(0)}_{\nu}= i\delta\overline{\delta}
 A^{(0)}_{\nu} .
\end{array}
\eqno{(3.1)}
$$

\noindent Then we study its relationship to the
renormalized version, and in order to facilitate
understanding of the nature of the problem we start
from its abelian version or QED, namely,

$$
\begin{array}{lll}
\partial^{\mu}F^{(0)}_{\mu\nu} +
e_0J^{(0)}_{\nu}=-\partial_{\nu}B^{(0)} ,
\end{array}
\eqno{(3.2)}
$$

\noindent The multiplicative renormalization of
fields and parameters relevant to this equation can
be summarized as

$$
\begin{array}{lll}
A_{\mu}^{(0)}= Z_3^{1/2}A_{\mu},& B^{(0)}=
 Z_3^{-1/2}B &J_{\nu}^{(0)}= J_{\nu},\\
\end{array}
\eqno{(3.3)}
$$

$$
\begin{array}{ll}
e_0= Z_3^{-1/2}e,& \alpha_0=
 Z_3\alpha .\\
\end{array}
\eqno{(3.4)}
$$

\noindent The renormalized version of Eq. (3.2) is
given by

$$
\begin{array}{lll}
\partial^{\mu}F_{\mu\nu} +
e\tilde{J}_{\nu}=-\partial_{\nu}B ,
\end{array}
\eqno{(3.5)}
$$

\noindent where

$$
\begin{array}{l}
e\tilde{J}_{\nu} = Z_3^{-1}\left[eJ_{\nu}+
(1-Z_3)\partial_{\nu}B\right] .
\end{array}
\eqno{(3.6)}
$$

\noindent It so happens that both $\partial^{\mu}F_{\mu\nu}$ and
$\partial_{\nu}B$ are multiplicatively renormalized but with
different multiplicative factors. This mismatch forces  to introduce
operator mixing for renormalization \cite{11}. Since both
$\partial^{\mu}F_{\mu\nu}$ and $\partial_{\nu}B$ are finite
operators so must be $\tilde{J}_{\nu}$ too.

Essentially the same situation takes place in QCD,
and the renormalized version of (3.1) is given by

$$
\begin{array}{lll}
\partial^{\mu}F^a_{\mu\nu} +
g\tilde{J}^a_{\nu}=i\delta\overline{\delta}A_{\nu}^a
,
\end{array}
\eqno{(3.7)}
$$

\noindent where $\tilde{J}^a_{\nu}$ is a linear
combination of $J^a_{\nu}$ and
$i\delta\overline{\delta}A_{\nu}^a$. The space
integral of $J^a_0$ gives the color charge $Q^a$.

$$
\begin{array}{l}
Q^a= \int d^3x J^a_0(x) ,
\end{array}
\eqno{(3.8)} $$

\noindent satisfying the commutation relations of
the color SU(3) algebra,

$$
\begin{array}{l}
\left[ Q^a, Q^b\right]= if_{abc}Q^c .\\
\end{array}
\eqno{(3.9)} $$

\noindent With the help of Eq. (3.7) we can write
down an equation for two-point Green's functions of
the form

$$
\begin{array}{lll}
\langle
\partial^{\lambda}F^a_{\lambda\mu}(x),A_{\nu}^b(y)\rangle +
\langle g\tilde{J}^a_{\mu}(x),A_{\nu}^b(y)\rangle
=\langle i\delta\overline{\delta}A_{\mu}^a
,A_{\nu}^b(y)\rangle .
\end{array}
\eqno{(3.10)}
$$

\noindent In what follows we shall study the
structure of the Fourier transforms of these
Green's functions.

Let $F_{\mu}$ and $G_{\nu}$ be vector fields and
introduce

$$
\begin{array}{l}
\langle F_{\mu}(x),G_{\nu}(y)\rangle
=\frac{-i}{(2\pi)^4}\int d^4k e^{ik\cdot(x-y)}
T_{\mu\nu}(k) ,
\end{array}
\eqno{(3.11)}
$$

\noindent and the Fourier transform of $\langle
F_{\mu},G_{\nu}\rangle$ is denoted by

$$
\begin{array}{lll}
T_{\mu\nu}(k) =\mbox{FT}\langle
F_{\mu},G_{\nu}\rangle .
\end{array}
\eqno{(3.12)}
$$

\noindent Then $T_{\mu\nu}$ can be expressed as a
linear combination of two covariants:

$$
\begin{array}{l}
T_{\mu\nu}(k)=
-\frac{k_{\mu}k_{\nu}}{k^2+i\epsilon}T_0(k^2)
-(g_{\mu\nu} - \frac{k_{\mu}
k_{\nu}}{k^2+i\epsilon}) T_1(k^2) .
\end{array}
\eqno{(3.13)} $$

 Next we introduce two conditions:

\noindent Condition 1):\\
\noindent Assume $\partial_{\mu}F_{\mu}=0$ and/or
$\partial_{\mu}G_{\mu}=0$, then we get

$$
\begin{array}{ll}
T_0(k^2)=T_0, & constant.
\\\end{array}\eqno{(3.14)}
$$

\noindent Condition 2):\\
Assume in addition to the condition 1) that
$Q\vert0\rangle=\langle 0\vert Q=0$, where

$$
\begin{array}{l}
Q=\int d^3x F_0(x),
\\\end{array}\eqno{(3.15)}
$$

\noindent then we  have

$$\begin{array}{l} T_0=0 .
\\\end{array}\eqno{(3.16)}
$$

\noindent Indeed, when this condition is not met we
have a broken symmetry and the Nambu-Goldstone
boson shows up in the form $T_0\neq 0$.

All three terms in Eq. (3.10) satisfy the condition
1) and their Lehmann representations are given as
follows. First, by taking account of the
antisymmetry between subscripts $\lambda$ and
$\mu$, we find

$$
\begin{array}{l}
\mbox{FT}\langle F_{\lambda\mu},A_{\nu}\rangle
=-i(k_{\lambda}g_{\mu\nu} - k_{\mu}g_{\lambda\nu})\left[
\frac{R}{k^2+i\epsilon} + \int dm^2 \frac{\sigma_1
(m^2)}{k^2-m^2+i\epsilon} \right] ,\\
\end{array}
\eqno{(3.17)} $$

\noindent so that we obtain

$$
\begin{array}{l}
\mbox{FT}\langle
\partial^{\lambda}F_{\lambda\mu},A_{\nu}\rangle =-R
\frac{k_{\mu}k_{\nu}}{k^2-i\epsilon} +
(k^2g_{\mu\nu}-k_{\mu}k_{\nu}) \int dm^2
\frac{\sigma_1 (m^2)}{k^2-m^2+i\epsilon}
.\\
\end{array}
\eqno{(3.18a)} $$

\noindent Then, thanks to the condition 2), we have in the absence
of operator mixing
$$
\begin{array}{l}
\mbox{FT}\langle g\tilde J_{\mu},A_{\nu}\rangle
=-(k^2g_{\mu\nu}-k_{\mu}k_{\nu}) \int dm^2
\frac{\sigma_2 (m^2)}{k^2-m^2+i\epsilon}  .\\
\end{array}
\eqno{(3.18b)} $$

\noindent Finally we have

$$\begin{array}{l}
\mbox{FT}\langle i\delta
\overline{\delta}A_{\mu},A_{\nu}\rangle  =-C
\frac{k_{\mu}k_{\nu}}{k^2+i\epsilon}
-(k^2g_{\mu\nu}-k_{\mu}k_{\nu}) \int dm^2
\frac{\sigma_3 (m^2)}{k^2-m^2+i\epsilon }
.\\
\end{array}
\eqno{(3.18c)}
$$

In what follows we shall study the properties of
these integral representations in more detail in
QED and QCD.

In QED we have $R=C=1$, so that

$$
\begin{array}{l}
\mbox{FT}\langle\partial^\lambda F_{\lambda\mu},A_{\nu}\rangle  =
-\frac{k_{\mu}k_{\nu}}{k^2+i\epsilon} +
(k^2g_{\mu\nu}-k_{\mu}k_{\nu}) \int dm^2
\frac{\sigma (m^2)}{k^2-m^2+i\epsilon}  ,\\
\end{array}
\eqno{(3.19)} $$

\noindent and

$$\begin{array}{l}
\mbox{FT}\langle -\partial_{\mu}B,A_{\nu}\rangle
= -\frac{k_{\mu}k_{\nu}}{k^2+i\epsilon}   ,\\
\end{array}
\eqno{(3.20)}
$$

\noindent so that we have
$$
\begin{array}{l}
\mbox{FT}\langle e\tilde{J}_{\mu},A_{\nu}\rangle
=(k^2g_{\mu\nu}-k_{\mu}k_{\nu}) \int dm^2
\frac{\sigma (m^2)}{k^2-m^2+i\epsilon}  .\\
\end{array}
\eqno{(3.21)} $$

\noindent The absence of the massless pole term in
(3.21) reflects the fact that the conservation of
charge is not broken.

In QCD let us assume that the condition for color
confinement (2.18) is satisfied, then we have

$$
\begin{array}{l}
\mbox{FT}\langle
\delta\overline{\delta}A_{\mu},A_{\nu}\rangle
=-(k^2g_{\mu\nu}-k_{\mu}k_{\nu}) \int dm^2
\frac{\sigma_3 (m^2)}{k^2-m^2+i\epsilon}  ,\\
\end{array}
\eqno{(3.22)} $$ which justifies the absence of the operator mixing
in (3.18b).
\noindent Since $\tilde{J}_{\mu}$ is a linear
combination of $J_{\mu}$ and $i\delta\overline{\delta}A_{\mu}$, we
can form a linear combination of (3.18b) and (3.22) to find

$$
\begin{array}{l}
\mbox{FT}\langle g\tilde{J}_{\mu},A_{\nu}\rangle
=-(k^2g_{\mu\nu}-k_{\mu}k_{\nu}) \int dm^2
\frac{\tilde{\sigma}_2 (m^2)}{k^2-m^2+i\epsilon}  ,\\
\end{array}
\eqno{(3.23)} $$

\noindent where $\tilde{\sigma}_2$ is a linear
combination of $\sigma_2$ and $\sigma_3$. Then
substituting (3.18a), (3.22) and (3.23) for Green's
functions in Eq. (3.10) we find

$$
\begin{array}{l}
R=0,\\
\end{array}
\eqno{(3.24)} $$

\noindent or

$$
\begin{array}{l}
\mbox{FT}\langle \partial^\lambda F_{\lambda\mu},A_{\nu}\rangle
=(k^2 g_{\mu\nu}-k_{\mu}k_{\nu}) \int dm^2
\frac{\sigma_1 (m^2)}{k^2-m^2+i\epsilon}  ,\\
\end{array}
\eqno{(3.25)} $$

\noindent Thus we find that there are no
contributions to this two-point function from
massless gluon states indicating the absence of the
massless gluon.

Then what would be the mechanism for generating the
finite gluon mass? A possible candidate is the
gluon condensate defined by

$$
\begin{array}{l}
\langle 0\vert A^a_{\mu}(x)A^b_{\mu}(x)\vert
0\rangle
=\delta_{ab}K\neq 0 .\\
\end{array}
\eqno{(3.26)} $$

\noindent A BRS invariant modification of (3.26) has been considered
by Kondo {\it et al} to introduce an effective gluon mass \cite{12}.
For phenomenological implications of such a gluon condensate as in
(3.26), see e.g. \cite{13}. In the Lagrangian density (2.2a) we find
a term quartic in the gauge field and replacement of a pair of gauge
fields by their vacuum expectation values (3.26) yields a mass term.
By keeping only bilinear terms relevant to the description of the
free massive gluon field, we find

$$
\begin{array}{l}
{\cal L}_g= \frac{1}{4} A_a^{\mu\nu}
A^a_{\mu\nu}-\frac{1}{2}m_0^2A_a^{\mu} A^a_{\mu} ,
\end{array}
\eqno{(3.27)}
$$

\noindent where

$$
\begin{array}{l}A^a_{\mu\nu}= \partial_{\mu} A^a_{\nu} -
\partial_{\nu} A^a_{\mu},
\end{array}
\eqno{(3.28)} $$

$$
\begin{array}{lll}
m_0^2 = g_0^2 c_2(G)K ,
\end{array}
\eqno{(3.29)} $$

\noindent where $c_2(SU(N))= N$. This gluon condensate does not
violate the color SU(3) symmetry nor the BRS invariance provided
that the mass term in (3.27) is dynamically generated.

\section{Equivalence Class of Gauges}

In Sec. 2 we have introduced the QCD Lagrangian
density that depends on the gauge parameter
$\alpha$. We shall study here how theoretical
predictions depend on this parameter.

In perturbation theory all the observable
quantities are independent of the choice of the
gauge parameter, but this is not the case in the
non-perturbative approach as we shall see later in
this section. Let us consider a class of Lagrangian
densities $\{{\cal L}_{\alpha}\}$ representing a
gauge theory such as QCD. Assume that all the
members of this set are BRS invariant,

$$
\begin{array}{lll}
\delta {\cal L}_{\alpha}=0 ,
\end{array}
\eqno{(4.1)} $$

\noindent and further that the difference between
any two elements of this set  is exact so that it
can be expressed as the BRS transform of a certain
operator ${\cal M}$,

 $$
\begin{array}{lll}
\Delta {\cal L}= {\cal L}_{II}-{\cal L}_{I}= \delta
{\cal M} ,
\end{array}
\eqno{(4.2)} $$

\noindent then this set $\{ {\cal L}_{\alpha}\}$ is
called an equivalence class of gauges (ECG) in a
loose sense. We shall show latter, however, that we
have to introduce an additional condition for
identifying the element of an ECG in the
non-perturbative approach.

Lagrangian densities corresponding to different
choices of $\alpha$ in (2.2) belong to the same
class in perturbation theory since we have

 $$
\begin{array}{lll}
\Delta {\cal L}= \frac{1}{2}(\Delta \alpha)B\cdot
B= - \frac{i}{2}(\Delta \alpha)
\delta(\overline{c}\cdot B)  ,
\end{array}
\eqno{(4.3)} $$

\noindent or

 $$
\begin{array}{lll}
{\cal M}=  - \frac{i}{2}(\Delta \alpha)
(\overline{c}\cdot B).
\end{array}
\eqno{(4.4)} $$

\noindent  Now we introduce Green's functions in two gauges of the
same ECG; then they are related to one another through the
Gell-Mann-Low relation \cite{14}:

$$
\begin{array}{lll}
\langle A(x_1)B(x_2)\cdots \rangle_{II} = \langle
A(x_1)B(x_2)\cdots exp(i\Delta S)\rangle_{I}  ,
\end{array}
\eqno{(4.5)} $$

\noindent where $A$, $B$, $\cdots$ are local
operators, and

$$
\begin{array}{l}
\Delta S = \int d^4x \Delta {\cal L} = \delta \int
d^4x  {\cal M} .
\end{array}
\eqno{(4.6)} $$

\noindent In particular, when all the local
operators are BRS invariant, namely

$$
\begin{array}{lll}
\delta A=\delta B=\cdots =0 ,
\end{array}
\eqno{(4.7)} $$

\noindent we obtain

$$
\begin{array}{lll}
\langle A(x_1)B(x_2)\cdots \rangle_{II} = \langle
A(x_1)B(x_2)\cdots \rangle_{I} ,
\end{array}
\eqno{(4.8)} $$

\noindent by expanding the r.h.s of Eq. (4.5) in
powers of $\Delta S$. Since $A$, $B$,$\cdots$ are
closed and $\Delta S$ is exact, we have

$$
\begin{array}{lll}
\langle A,B,\cdots,(\Delta S)^n \rangle = 0 .
\end{array}
\eqno{(4.9)} $$

\noindent Since we are exploiting a series
expansion in powers of $\Delta S$, the proof of
(4.8) is based on an assumed convergence of this
series.

In what follows we shall restrict ourselves to the
Lagrangian density (2.2) and assume that all the
quark masses have certain fixed values. Then a
theory is characterized by $\alpha$ and $g^2$.  We
shall plot $(\alpha,g^2)$ on a two-dimensional
half-plane

$$
\begin{array}{lll}
 -\infty<\alpha<\infty,&g^2>0 ,
\end{array}
\eqno{(4.10)} $$

\noindent then each point on this half-plane corresponds to one
theory. Now we introduce renormalization group (RG) with the
generator of this RG given by \cite{10}

$$
\begin{array}{lll}
{\cal D} = \mu \frac{\partial}{\partial \mu} +
\beta (g) \frac{\partial}{\partial g} - 2 \alpha
\gamma_{V} (g,\alpha) \frac{\partial}{\partial
\alpha},
\end{array}
\eqno{(4.11)} $$

\noindent where $\mu$ denotes the renormalization point with
dimension of mass and $\gamma _{V}$ is the anomalous dimension of
the color gauge field. Since we have discussed applications of RG to
Green's functions elsewhere \cite{3,10}, we shall confine ourselves
to the study of running parameters defined by

$$
\begin{array}{l}
\overline{g}(\rho)=  exp(\rho {\cal D}) \cdot g , \\
\overline{\alpha}(\rho)=  exp(\rho {\cal D}) \cdot \alpha , \\
\overline{\mu}(\rho)=  exp(\rho {\cal D}) \cdot \mu =e^{\rho}\mu , \\
\end{array}
\eqno{(4.12)} $$

\noindent where $\rho$ denotes the parameter of RG.
In what follows we shall concentrate our attention
on $\overline{g}(\rho)$ and
$\overline{\alpha}(\rho)$.

When we increase $\rho$ from 0 to $\infty$, the
point$(\overline{\alpha}(\rho), \overline{g}(\rho))$ moves along a
line called the RG flow line (RGFL). Two theories corresponding to
two points on the same RGFL are completely equivalent and physically
identical as has been clarified in the applications of RG to Green's
functions \cite{3,10}. The asymptotic limits of these running
parameters are denoted by

$$
\begin{array}{ll}
\bar{g} (\infty) = g_{\infty},& \bar{\alpha}
(\infty) = \alpha_{\infty},
\end{array}
\eqno{(4.13)} $$

\noindent and such a point represents a sink of the
RGFL. Asymptotic freedom that has been assumed
throughout this article is characterized by

$$
\begin{array}{l}
g_{\infty} = 0.
\end{array}
\eqno{(4.14)} $$

\noindent It should be emphasized that
$\alpha_{\infty}$ can assume only three alternative
values depending on the choice of the starting
point of the RGFL, namely

$$\begin{array}{lll} \alpha_{\infty}= -\infty, & 0 , &
\alpha_0 ,
\end{array}
\eqno{(4.15)} $$

\noindent where $\alpha_0$ depends only on the
number of quark flavors $N_f$ in QCD

$$
\begin{array}{l}
\alpha_{0} = - \frac{1}{3} (13- \frac{4}{3} N_f).\\
\end{array}
\eqno{(4.16)}$$

\noindent Without loss of generality we shall
restrict ourselves to the following case:

$$\begin{array}{ll}
\alpha_0 > 0, &\mbox{or} ~  N_f < 10 .\\
\end{array}
\eqno{(4.17)}$$

\noindent The value of $\alpha_{\infty}$ depends only on the initial
value of $\alpha$, namely:

$$\begin{array}{lll} \alpha_{\infty} = \left \{
\begin{array}{cc}
 \alpha_0, & \mbox{for}~  \alpha > 0,\\
 0, & \mbox{for}~ \alpha = 0, \\
-\infty,& \mbox{for}~  \alpha < 0.
\end{array}
\right .
\end{array}
\eqno{(4.18)}
$$

We recall the arguments on the ECG. Equivalence of
two gauges characterized by  $(\alpha_1,g_1^2)$ and
$(\alpha_2,g_2^2)$ was based on series expansion in
powers of $\Delta \alpha$, and when they belong to
the same ECG so do also
$(\overline{\alpha}_1,\overline{g}_1^2)$ and
$(\overline{\alpha}_2,\overline{g}_2^2)$. The
equivalence of the latter is based on the series
expansion in powers of
$\Delta\overline{\alpha}(\rho)$. Now assume that
$\alpha_1<0$ and $\alpha_2>0$; however small the
difference $\Delta \alpha=\alpha_2-\alpha_1$ might
be, $\alpha_1$ tends to $\infty$ and $\alpha_2$ to
$\alpha_0$ so that $\Delta \overline{\alpha}(\rho)$
tends to $\infty$. This fact casts a doubt on the
convergence of the power series and we may conclude
that these two points do not belong to the same
ECG. Thus it is very likely that we have two sets
of ECG defined on the two-dimensional parameter
half-plane

$$\begin{array}{l}
{\cal D} (-\infty )  =  \{ (\alpha,g^2): \alpha<0,
g^2>0\},\\
\end{array}
\eqno{(4.19)}
$$

$$\begin{array}{l}
{\cal D} (\alpha_0)  =  \{ (\alpha,g^2): \alpha>0,
g^2>0\} .\\
\end{array}
\eqno{(4.20)}
$$

\noindent Both of them are two-dimensional domains,
but in addition we have a one-dimensional line
which forms the border between them, namely,

$$\begin{array}{l}
L(0)   =  \{ (\alpha,g^2): \alpha=0, g^2>0\} .\\
\end{array}
\eqno{(4.21)}
$$

\noindent Indeed, $\bar\alpha(\rho)$ shows a discontinuity at
$\alpha=0$ when $\rho$ is chosen sufficiently large as it is the
case for $\alpha_\infty$. Then with the help of Eq. (4.9) of [4],
i.e.

$$
G(p_i; g,\alpha,\mu)= \exp\left[\int_0^\rho
d\rho'\bar\gamma(\rho')\right]\cdot G(p_i;\bar g(\rho),
\bar\alpha(\rho),\bar\mu(\rho))\ ,
$$
where $\gamma$ denotes the anomalous dimension of the Green's
function in question, we may conclude that Green's functions also
develop a discontinuity at $\alpha=0$. This discontinuity is
certainly reflected in the Green's functions, for instance, in the
form of the residue of the massless pole of the gluon propagator
studied in the preceding section.

In this connection it is also important to recognize that particle
masses defined as the pole positions of propagators are independent
of $\rho$ and thus are RG invariant as is clear from the above
relationship.

Thus it seems likely that we have three ECG specified by the value
of $\alpha_{\infty}$.

It has been shown already that the renormalization constant $Z_3$ of
the color gauge field is given by \cite{3,10}

$$\begin{array}{l}
Z_3^{-1}   = \frac{\alpha}{\alpha_{\infty}} .\\
\end{array}
\eqno{(4.22)}
$$

\noindent This constant is related to $C$ in (2.19) through the
formula \cite{1}-\cite{4}

$$\begin{array}{lll} \overline{C} (\rho) &=& \bar{a}
(\rho) - 2 \int_{\rho}^{\infty} d\rho ^{'}
(\bar{\gamma} _{V} (\rho^{'}) + \bar{\gamma} _{FP}
(\rho^{'})) \bar{a} (\rho^{'})\\ && \times \exp
\left[ -2 \int_{\rho}^{\rho^{'}} d \rho^{''}
\bar{\gamma}_{FP} (\rho^{''})\right],\\
\end{array}
\eqno{(4.23)} $$

\noindent where $\gamma_{FD}$ is the anomalous
dimension of the Faddeev Popov ghost fields and
$\overline{a}(\rho)$ is given by

$$\begin{array}{l}
 \overline{a}(\rho)=
 \frac{\overline{\alpha}(\rho)}{\alpha_{\infty}}.\\
\end{array}
\eqno{(4.24)} $$

\noindent The validity of Eq. (4.23) has been tested in QED with the
result that it reproduces $C=1$ correctly and the lack of asymptotic
freedom in QED does not allow $C$ to vanish. Hence charge
confinement does not take place in QED.
 Therefore we have

$$
\begin{array}{l}
 Z_3^{-1}= \mbox{{\Large $\{$}} \begin{array}{ll}
 0, & \mbox{in}~ {\cal D}(-\infty) ~ \mbox{and}  ~L(0),\\
\frac{\alpha}{\alpha_0}, & \mbox{in}~ {\cal
D}(\alpha_0) .\\
\end{array}\\
\end{array}
\eqno{(4.25)} $$

\noindent Combining (4.23) and (4.25) we find

$$\begin{array}{ll}
C= 0, & \mbox{in}~ {\cal D}(-\infty) ~ \mbox{and}  ~L(0),\\
C \neq 0, & \mbox{in}~ {\cal D}(\alpha_0) .\\
\end{array}
\eqno{(4.26)} $$

\noindent This amounts to the conclusion that gluons are massive in
the confinement phase, ${\cal D}(-\infty)$ and $L(0)$, whereas they
remain massless in the deconfinement phase ${\cal D}(\alpha_0)$.

The case of $\alpha_0<0$ is slightly more
complicated, but essential features such as the
existence of two phases are the same.

\section{Conclusions}

Theories described by the Lagrangian density (2.1)
corresponding to different choices of the gauge
parameter $\alpha$ belong to the same ECG in
perturbation theory. This class is split into
three, however, in non-perturbative approach
depending on their convergence properties. Green's
functions in QCD are expanded in double power
series in $g^2$ and $\alpha g^2$, and the critical
parameters governing the convergence of the
expansion are the asymptotic values of $g^2$,
$\alpha$, and $\alpha g^2$. The last one cannot be
evaluated exactly, but is is still possible to
judge whether it is equal to zero or non-zero.

In order to clarify the differences among the three
classes in the convergence properties we shall
tabulate them in what follows:

Case 1): $\alpha_0>0$ or $N_f<10$.

\begin{center}
 \vl\begin{tabular}{||l|l|l|l|l||}\hline\hline
region& $(\alpha g^2)_{\infty}$&  $Z_3^{-1}$&confinement &gluon mass\\
\hline\cline{1-5}

$D(-\infty)$: & $\neq 0$     & 0 &yes     &$\neq
0$\\ \hline

$D(0)$ or $L(0)$:& 0 & 0 &yes &$\neq 0$\\\hline

$D(\alpha_0)$ & 0   & $\neq 0,\infty$      &no      &0\\
\hline \hline
\end{tabular}\vl

\end{center}

In this case $D(0)$ or $L(0)$ is a straight line
$\alpha=0$ and forms the border between
$D(-\infty)$ and $D(\alpha_0)$.

Case 2): $\alpha_0<0$ or $10\leq N_f\leq 16$.

\begin{center}
 \vl\begin{tabular}{||l|l|l|l|l||}\hline\hline
region& $(\alpha g^2)_{\infty}$&  $Z_3^{-1}$&confinement &gluon mass\\
\hline\cline{1-5}

 $D(-\infty)$: & $\neq 0$     & 0 &yes     &$\neq 0$\\ \hline

$D(\alpha_0)$ or $L(\alpha_0)$:& 0 &$\neq 0,\infty$
&no &$ 0$\\\hline

$D(0)$ & 0   & $\infty$      &no      &0\\
\hline \hline
\end{tabular}\vl

\end{center}

In this case $D(\alpha_0)$ or $L(\alpha_0)$ is a
RGFL and forms  the border between $D(-\infty)$ and
$D(0)$.

Our naive belief in the independence of physics on the gauge
parameter is no longer justified, but instead three phases show up.
There are confinement and deconfinement phases, and in the former
gluons turn out to be massive and in the latter they remain
massless.

Color confinement is one of the central issues in QCD so that there
are various interpretations of this feature. In this paper we have
adopted the interpretation that colored particles are not subject to
observation just because colored states are unphysical in the sense
of Eq. (2.16). Then the emergence of massive gluons is an inevitable
consequence of this interpretation.

\vskip .5cm

{\bf Acknowledgements:} We are grateful to Paul Hoyer, Archil
Kobakhidze and Anca Tureanu for useful discussions on the
phenomenological situation with massive gluons. One of the authors
(K.N.) is partially supported by Grant-in-Aid for Scientific
Research from the Ministry of Education, Culture, Sports, Science
and Technology of Japan. The financial support of the Academy of
Finland under the projects no. 54023 and 104368 is greatly
acknowledged.


\begin{thebibliography}{99}

\bibitem{1}
K. Nishijima, Int. J. Mod. Phys. {\bf A9}, 3799 (1994).

\bibitem{2}
K. Nishijima, Int. J. Mod. Phys. {\bf A10}, 3155 (1995).


\bibitem{3}
K. Nishijima, Czech. J. Phys. {\bf 46}, 1 (1996).

\bibitem{4}
M. Chaichian and K. Nishijima, Eur. Phys. J. {\bf C22}, 463 (2001).


\bibitem{5}
D. J. Gross and F. Wilczek,  Phys. Rev. Lett. {\bf 30}, 1343 (1973).

\bibitem{6}
H. D. Politzer,  Phys. Rev. Lett.  {\bf 30}, 1346 (1973).

\bibitem{7}
C. Becchi, A. Rouet and R. Stora, Ann. Phys.   {\bf 98}, 287 (1976).


\bibitem{8}
K. Nishijima, Prog. Theor. Phys. {\bf 80}, 897 (1988).

\bibitem{9}
K. Nishijima, Prog. Theor. Phys. {\bf 80}, 905 (1988).


\bibitem{10}
K. Nishijima and N. Takase, Int. J. Mod. Phys. {\bf A11}, 2281
(1996).

\bibitem{11'}
R. Oehme and W. Zimmermann, Phys. Rev. {\bf D 21}, 471, 1661 (1980).

\bibitem{11}
K. Nishijima,  Phys. Rev. {\bf 119}, 485 (1960).



\bibitem{12} K.-I. Kondo, Phys. Lett. {\bf B 619}, 377 (2005), hep-th/0504088;
Phys. Lett. {\bf B 572}, 210 (2003), hep-th/0306195 and references
therein.

\bibitem{13} F.V. Gubarev, L. Stodolsky and V. I. Zakharov,  Phys. Rev. Lett. {\bf 86}, 2220
(2001),
hep-ph/0010057; \\
F.V. Gubarev and V. I. Zakharov, Phys. Lett. {\bf B 501}, 28 (2001),
hep-ph/0010096.


\bibitem{14}
M. Gell-Mann and F. Low, Phys. Rev. {\bf 84 }, 350
(1951).







\end{thebibliography}
\end{document}